\documentclass[twocolumn,english,prx]{revtex4-1}
\usepackage{color}
\usepackage[utf8]{inputenc}
\usepackage{graphicx}
\usepackage[T1]{fontenc}
\usepackage{float}
\usepackage{natbib}
\usepackage{textcomp}
\usepackage{amsmath}
\usepackage{amssymb}
\usepackage{graphicx}
\usepackage{esint}
\usepackage{natbib}
\usepackage{xcolor}
\usepackage{multirow}
\usepackage{ulem}

\begin{document}
\title{Exciton localization on p-i-n junctions in two-dimensional crystals}

\author{Bart\l{}omiej Szafran}

\affiliation{AGH University of Science and Technology, Faculty of Physics and
Applied Computer Science,\\
 al. Mickiewicza 30, 30-059 Kraków, Poland}

\begin{abstract}
We consider a neutral exciton localized on a model p-i-n junction
defined in a two-dimensional crystal: MoSe$_2$ and phosphorene,
using a variational approach to the effective mass Hamiltonian. 
The variational solution to the problem with non-separable center of mass 
provides the exciton density in the 
real space and accounts for the kinetic energy due to the exciton localization.
 For low values of the potential step across the junction,
the exciton occupies an area which is much larger than the nominal width of the junction.
%and its energy remains essentially insensitive to the value of the step. 
Localization of the exciton within the junction area is accompanied by
the appearance of the dipole moment induced by the local electric field. 
The induced dipole moment becomes a linear function of the potential step only
when the step is sufficiently large. In consequence, the energy dependence
on the step value is non-parabolic. We demonstrate that the exciton
gets localized not exactly at the center of the junction but on the side which is
more energetically favourable for the heavier carrier: electron or hole.
\end{abstract}
\maketitle
\section{Introduction}

Reduction of dimensionality and large carrier effective masses in monolayer transition-metal
dichalcogenide \cite{tmdc1,tmdc3} and phosphorene \cite{xia,castelano} result in large exciton binding energies \cite{tran,seixas,kep3,varga,starkchav,starkchav2,peeters,rodin}
exceeding by two orders of magnitude the corresponding values in bulk semiconductors. 
Large binding energy allows for tuning the exciton line in the luminescent spectrum by the Stark effect
induced by an in-plane electric field \cite{starkchav} which leads to a red-shift of the electron-hole recombination energy 
due to the interaction of the intrinsic or induced exciton dipole moment with the field. 
The Stark effect was studied for states in quantum wells \cite{qw0,qw1} or quantum dots \cite{qd0,qd1,qd2,qd3} where 
the quantum confinement stabilized
the weakly bound exciton against dissociation by the electric field. 
Conversely, the Stark effect for strongly bound excitons in a transition metal
dichalcogenide has recently been used to trap the excitons on the electric field
at the p-i-n junction \cite{na}.
% The junction potential that separates the carriers' charges to the n or p regions
%confines the neutral excitons in the electric field locally induced in the intrinsic region \cite{na}.

The purpose of the present work is to describe the exciton confinement in a model of the p-i-n junction 
by a numerical albeit exact solution of the two-particle problem.
The  trapping mechanism was explained \cite{na} as due to an effective potential for the exciton that
contains a term due to the electric field $V_{eff}(x)=-\frac{1}{2} \alpha_p |F_x(x)|^2$, where $F_x$
is the electric field at the junction and $\alpha_p$ is the polarizability of the exciton, which for
the homogeneous electric field can be determined by the
solution of the electron-hole problem in the relative electron-hole coordinates \cite{na,starkchav,starkchav2}. 
The electric field in the junction \cite{na} is not homogeneous and the motion of the center
of mass of the exciton does not separate from the eigenproblem in the relative coordinates.
Therefore, the center of mass requires treatment on equal rights with the relative electron-hole coordinates
which we provide in this work. The solution accounts
for the kinetic energy due to the exciton confinement at the junction and provides
information on the exciton localization: its mean position and extent of its wave function in the real space.
We find that the range of exciton confinement depends strongly on the value of the potential
step at the junction, and for a low value of the step the confinement range can largely exceed the nominal width
of the junction. Then, the wave function is present in a region with a widely changing electric field
that makes the homogeneous field solution not directly adequate to a local field case. 
In particular, the electric field at the center of the junction is a linear function of the potential step
at the junction, but the induced dipole moment is not and the reaction of the exciton energy on the step is not parabolic as 
for Stark effect in the homogenous electric field.

\section{Theory}
We work with the effective band Hamiltonian for the electron-hole pair
at the junction
\begin{eqnarray}H&=&-\frac{\hbar^2}{2m^x_e}\frac{\partial^2}{\partial x_e^2}-\frac{\hbar^2}{2m^y_e}\frac{\partial^2}{\partial y_e^2}
-\frac{\hbar^2}{2m^x_h}\frac{\partial^2}{\partial x_h^2}-\frac{\hbar^2}{2m^y_h}\frac{\partial^2}{\partial y_h^2}\nonumber \\&
+&V(x_e)-V(x_h)-V_{eh}(r_{eh}),\end{eqnarray}
with $m^{y/x}_{e/h}$ standing for the electron or hole mass along the $x$ or $y$ directions.
We use the model potential of the p-i-n junction in the form  $V(x)=-e\frac{2V_0}{\pi}\arctan(\frac{x}{d})$ 
where $V_0$ defines the scale of the potential step along the junction and the variation rate
is given by $d$. The model potential is plotted in Fig. 1 for the electron ($V(x)$, red line) and the hole ($-V(x)$, black line).
The potential pushes the non-interacting carriers away from the junction center. 
In Eq. (1) $V_{eh}(r_{eh})$ is the effective 2D interaction
given by the Keldysh potential \cite{kep1,na,kep2,kep3,varga,ansatz,starkchav,starkchav2,peeters,rodin}, \begin{equation}
V_{eh}(r)=\frac{e^2}{4\pi \epsilon_0}\frac{\pi}{2\epsilon r_0}\left[H_0(r/r_0)-Y_0(r/r_0)\right],\end{equation}
where $H_0$ and $Y_0$ are the Struve and Bessel functions of the second kind,
$r_0$ is the screening length that is a measure of the polarizability of the 2D semiconductor
and $\epsilon$ is the dielectric constant. 
At large electron-hole distances $r>>r_0$ the interaction potential tends to the 3D Coulomb form $1/r$. At small electron-hole distances $r<<r_0$,
$V_{eh}$ acquires a logarithmic singularity of the 2D Coulomb potential. 
For evaluation of the interaction potential we use the method of Ref. \cite{ansatz}.
The intrinsic area where the exciton confinement occurs in the experiment \cite{na} has a length of several dozens of nanometers.
Here we discuss the results with the parameter $d$ ranging from 15 to 60nm. 
The variation of the potential by $V_0$ and $1.5V_0$ occurs at lengths of $2d$ and $2.4d$, respectively.
In the experimental situation \cite{na} the carriers are additionally confined by the space-charge density
outside the intrinsic region. Here we assume that the depletion region is arbitrarily wide for a description
of the exciton binding purely on the local electric field. 

Using the center-of-mass coordinates $X=\frac{m_e^xx_e+m_h^xx_h}{M_x}$, $Y=\frac{m^y_ey_e+m^y_hy_h}{M_y}$,
with $M_x=m_e^x+m_h^x$, $M_y=m_e^y+m_h^y$
and the relative electron-hole coordinates $x_{eh}=x_e-x_h$ and $y_{eh}=y_e-y_h$, the Hamiltonian 
is written as
\begin{eqnarray}H&=&-\frac{\hbar^2}{2M_x}\frac{\partial^2}{\partial X^2}
-\frac{\hbar^2}{2M_y}\frac{\partial^2}{\partial Y^2}-\frac{\hbar^2}{2\mu_x}\frac{\partial^2}{\partial x_{eh}^2}-\frac{\hbar^2}{2\mu_y}\frac{\partial^2}{\partial y_{eh}^2}\nonumber \\
&+&V(x_e)-V(x_h)-V_{eh}(r_{eh}),\end{eqnarray}
with the reduces masses $\mu_x=m_e^xm_h^x/(m_e^x+m_h^x)$ and $\mu_y=m_e^ym_h^y/(m_e^y+m_h^y)$.
The motion of the center of mass  along the junction separates from the rest of the coordinates, 
and we assume that the wave function over $Y$ has the form of a plane wave with a zero wave vector.
For a constant electric field oriented along the $x$ direction, studied for the discussion of
the Stark effect \cite{starkchav,starkchav2,na}, the center of mass coordinate $X$ also separates, but this is not the case for the present
problem with the non-linear potential of the junction $V$. 

The exciton states are determined using a variational approach \cite{varga} with Gaussian 
basis
\begin{eqnarray}
&&\Phi(x_{eh},y_{eh},X)  
 =\sum_{ijk} c_{ijk} \phi_{ijk}(x_{eh},y_{eh},X)= \\
&&\sum_{ijk} c_{ijk} \exp\left(-\frac{(x_{eh}-\kappa_i)^2}{\alpha}-\frac{(y_{eh}-\gamma_j)^2}{\beta}-\frac{(X-\eta_k)^2}{\gamma}\right)\nonumber,
\end{eqnarray}
where $c_{ijk}$ are the linear variational coordinates determined by solving the generalized eigenvalue problem
${\bf H}{\bf c}=E{\bf S}{\bf c}$, with the matrix elements $H_{i'j'k',ijk}=\langle \phi_{i'j'k'}|H|\phi_{ijk}\rangle$ 
and $S_{i'j'k',ijk}=\langle \phi_{i'j'k'}|\phi_{ijk}\rangle$.
The centers of the Gaussian are distributed on a 3D mesh with $\kappa_i=i\Delta_{x_{eh}}$,
$\gamma_j=j\Delta_{y_{eh}}$ and $\eta_k=k\Delta_{{X}}$, with $i$, $j$ and $k$ being integers ranging
from $-N$ do $N$. The values of $\Delta$ select the region in space to be described by the variational wave function.
The spacings $\Delta$'s and the localization parameters of the Gaussian 
$\alpha,\beta$ and $\gamma$ are determined as nonlinear variational parameters by minimization of the energy estimate.
%The optimal values of $\alpha\simeq \Delta X^2/1.5$, and similarly for the other set of parameters, which allows the overlaping Gaussian to 
%shape the wave function in the meaning of the variational principle.
Separate parameters for $x_{eh}$ and $y_{eh}$ are needed for phosphorene with its strongly anisotropic effective masses \cite{jakies,34} but also for the case of isotropic effective masses applied for MoSe$_2$ due to the external potential acting only in the $x$ direction.
The following results  are obtained for $N=8$, i.e. 17 Gaussians describing the wave function in each coordinate, i.e. for the total number of $17^3=4913$ Gaussian functions used in the basis.
The basis (4) is flexible enough to account for both the bound excitons and dissociated electron-hole pairs.

\begin{figure}[htbp]
\begin{tabular}{l}
\includegraphics[clip, height=0.1975\textwidth]{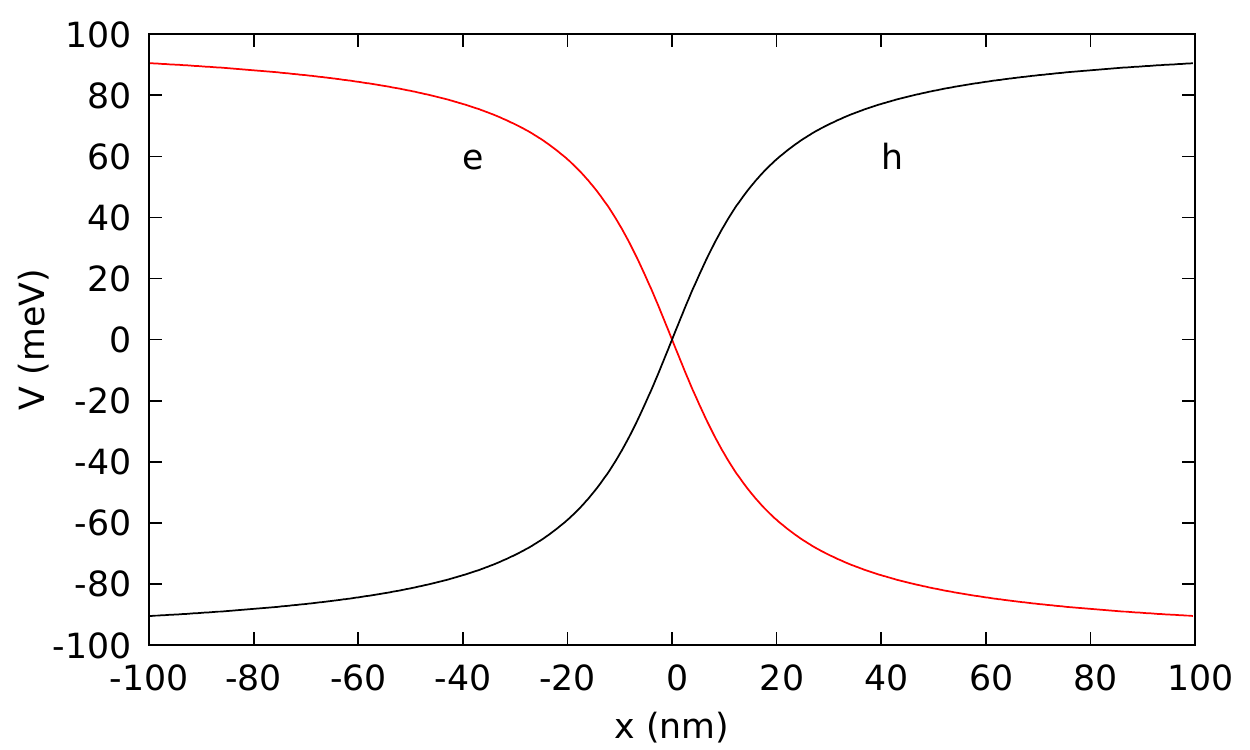} \put(-30,45){\color{black}(a)}  \\
\hspace*{-.2cm}\includegraphics[clip, height=0.2\textwidth]{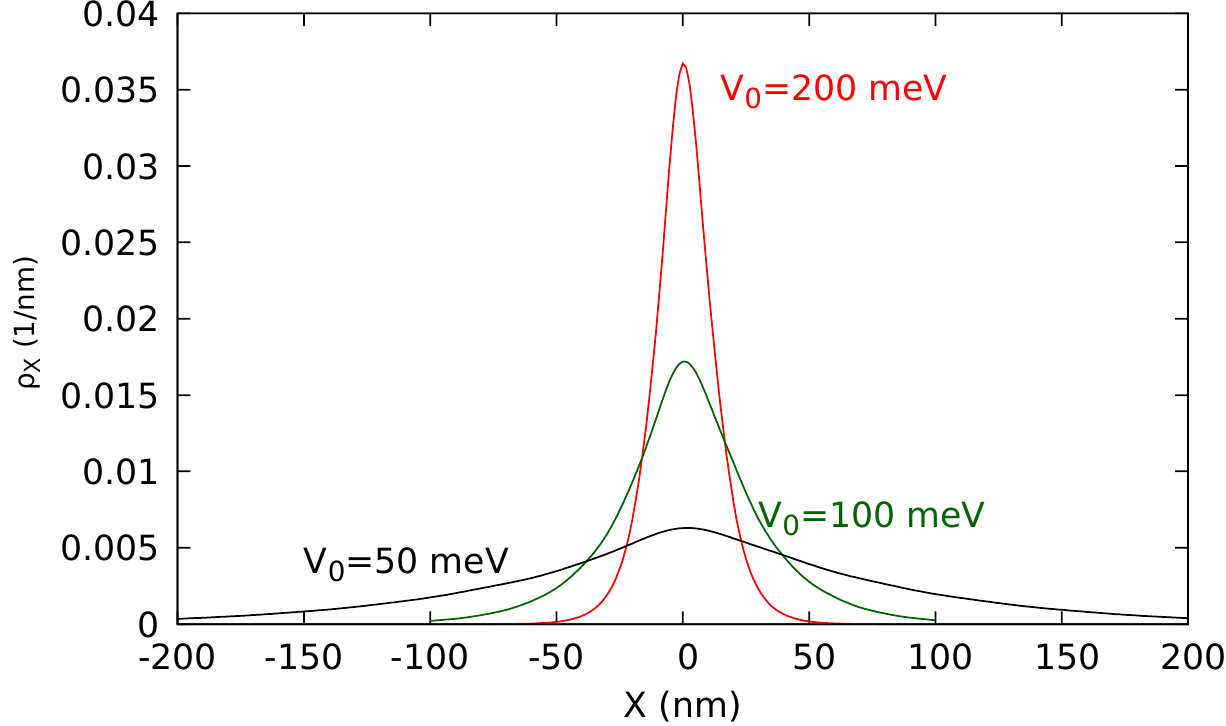} \put(-20,45){\color{black}(b)}  \\
\includegraphics[clip, height=0.187\textwidth]{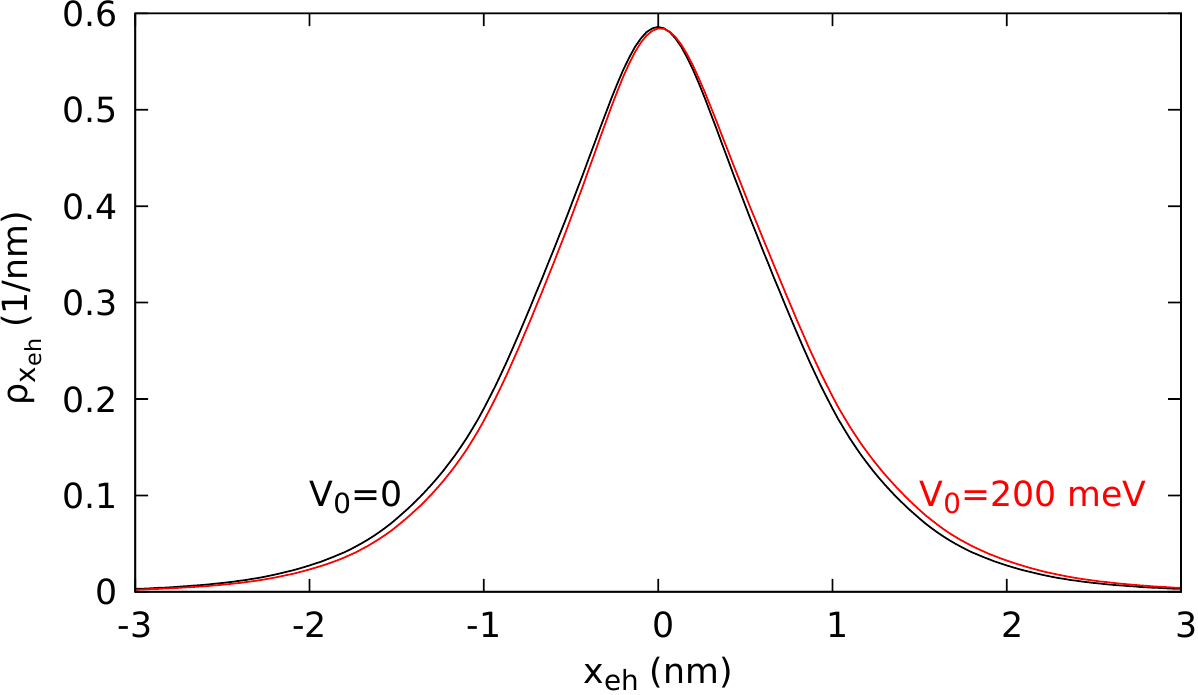} \put(-20,45){\color{black}(c)}  \\
\end{tabular}
\caption{
(a) The model potential of the p-i-n junction $V(x)=-\frac{2eV_0}{\pi} \arctan(\frac{x}{d})$ for the electron (red curve)
and $-V(x)$ potential for the hole (black curve) for $d=15$ nm and $V_0=100$ meV. (b) The exciton density as a function
of the center of mass position for $d=15$ nm and three values of $V_0$ for the lowest-energy bound exciton state. (c) The probability density as a function of the electron-hole
distance in the $x$ direction $x_{eh}$ for $V_0=0$ (black line) and $V_0=200$ meV (red line).
%In (b) and (c) results for $V_0=0,50$ and 100 meV correspond to the ground-state of the system and the ones for $V_0=200$ meV
%for the lowest-energy bound electron-hole pair. 
MoSe$_2$ parameters are applied.}
\label{wf}
\end{figure}

\begin{figure}[htbp]
\centering
%trim=left botm right top
\begin{tabular}{l}
\includegraphics[width=0.3\textwidth]{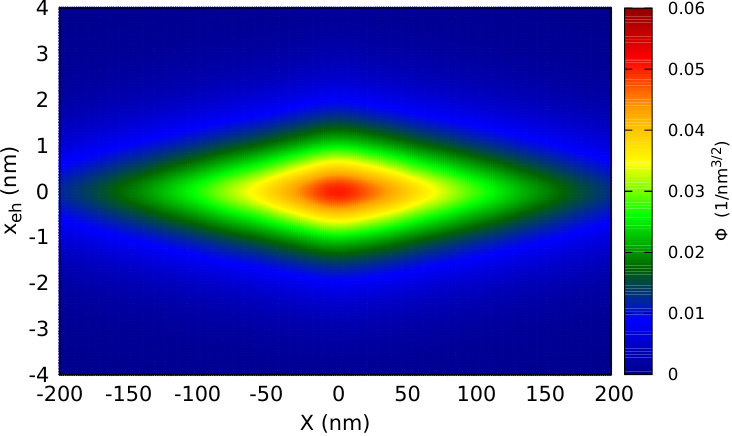} \put(-40,15){\color{white}(a)} \put(-120,15){\color{white}$V_0=50$ meV}  \\
\includegraphics[width=0.3\textwidth]{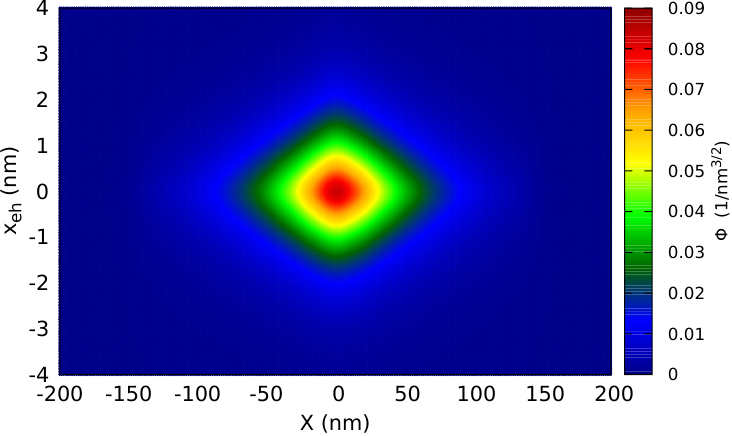} \put(-40,15){\color{white}(b)} \put(-120,15){\color{white}$V_0$=100 meV}  \\
\includegraphics[width=0.3\textwidth]{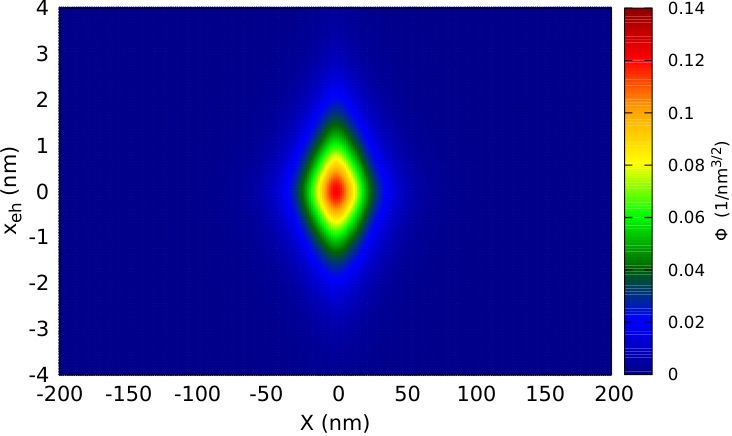} \put(-40,15){\color{white}(c)}  \put(-120,15){\color{white}$V_0$=200 meV} \\

\end{tabular}
\caption{
Cross section of the lowest-energy bound wave function taken at $y_{eh}=0$ $\Phi(x_{eh},y_{eh}=0,X)$ for $d=15$ nm and MoSe$_2$ parameters.
%The results of (a) and (b) shows the results for the ground-state and the ones in (c) the lowest-energy bound exciton state.
}
\label{wf}
\end{figure}
 
\begin{figure}[htbp]
\centering
%trim=left botm right top
\begin{tabular}{l}
\includegraphics[width=0.3\textwidth]{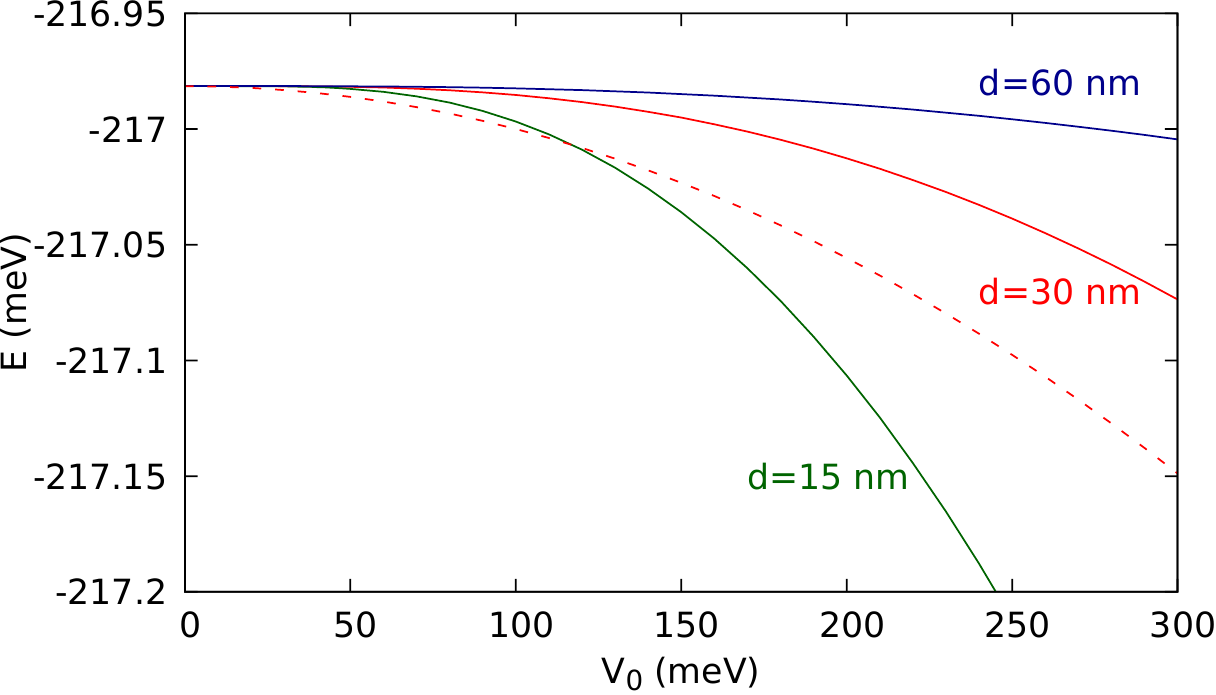} \put(-20,15){\color{black}(a)}  \\
\includegraphics[width=0.3\textwidth]{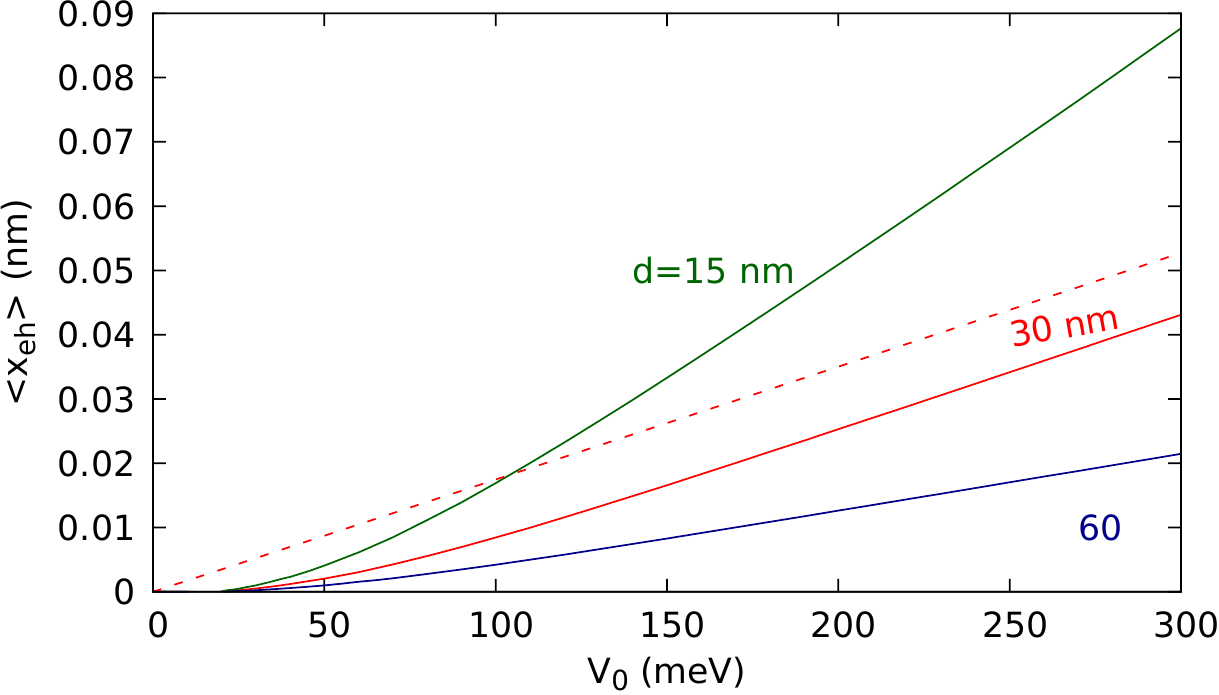} \put(-20,35){\color{black}(b)}  \\
\includegraphics[width=0.3\textwidth]{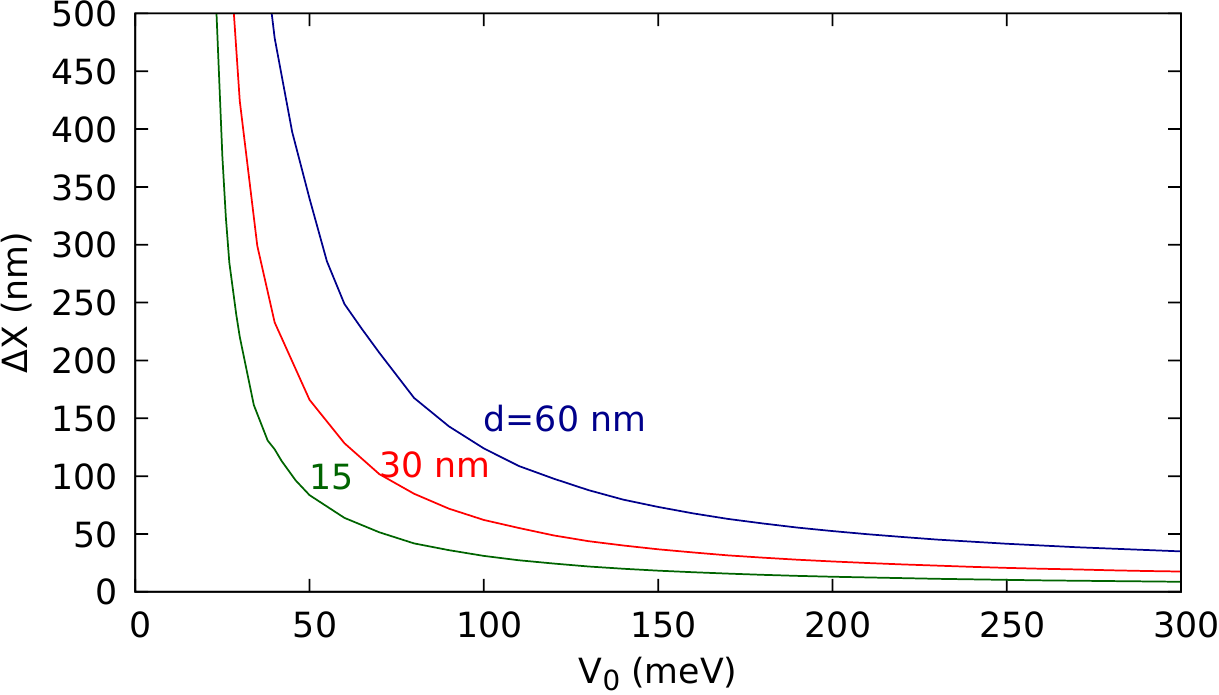} \put(-20,35){\color{black}(c)}  \\
\includegraphics[width=0.3\textwidth]{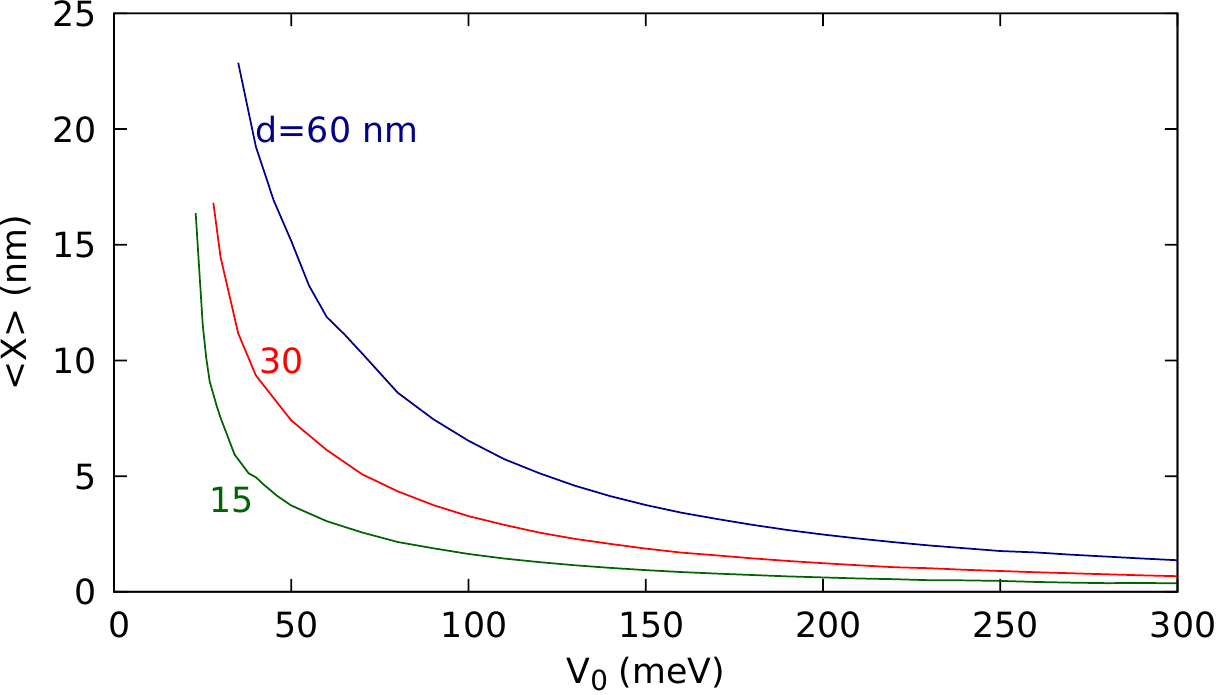} \put(-20,35){\color{black}(d)}  \\
\end{tabular}
\caption{
(a) Energy of the lowest bound exciton state as a function of $V_0$ for varied values of the width of the junction $d$
for MoSe$_2$. The dashed line shows the results for the junction potential replaced by its linear approximation
$V_l(x)=-e\frac{2V_0}{\pi}\frac{x}{d}$ for $d=30$ nm.
(b) The average shift of the electron and hole position along the direction of the junction $x_{eh}$.
The dashed line corresponds to the linear approximation as in (a).  (c) The exciton localization width calculated from the wave function as $\Delta X=\langle (X-\langle X \rangle )^2\rangle^{1/2}$.   
(d) The average position of the exciton center of mass. The line is plotted for $V_0$ where $\Delta X$ falls below 500 nm. 
%The  results presented in (a-d) correspond to the Hamiltonian ground state for $V_0<216.89/2$ meV. For larger $V_0$ 
The results correspond
to the lowest-energy bound exciton state.
}
\label{wf}
\end{figure}

\begin{figure}[htbp]
\centering
%trim=left botm right top
\includegraphics[width=0.4\textwidth]{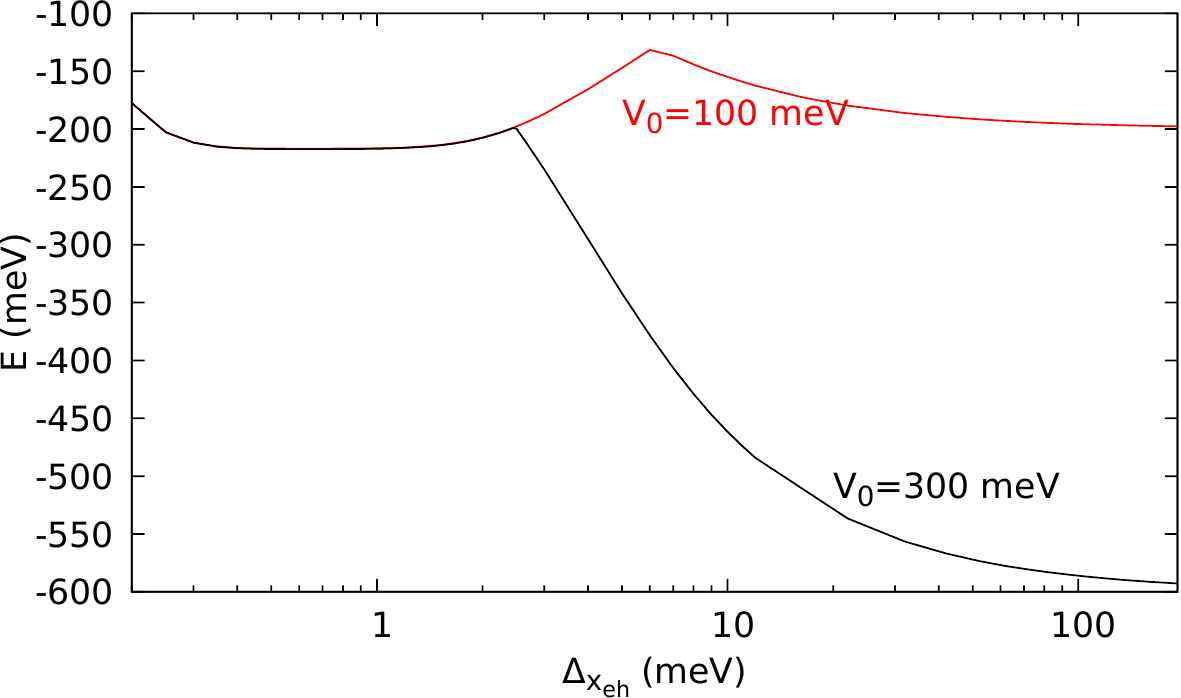}  \put(-155,80){\color{blue}$\uparrow$bound X}\put(-70,70){\color{blue} dissociated X$\rightarrow$ }
\caption{
The lowest energy estimate for $d=15$ nm with $V_0=100$ meV (red line) and $300$ meV (black line) for fixed $\Delta_X=18$ nm (see the description of the variational wave function given by Eq. (4)), as a function of  $\Delta_{x_{eh}}$.
For this plot we keep $\Delta_{y_{eh}}=\Delta_{x_{eh}}$ and the Gaussian localization parameters 
 $\alpha=\frac{\Delta_X^2}{1.5}$, and $\beta=\gamma=\frac{\Delta x_{eh}^2}{1.5}$ 
The minimum near $\Delta_{x{eh}}=0.8$ nm corresponds to the  bound exciton.
The results for large $\Delta_{x{eh}}$ limit tend to $-2V_0$ and correspond to the dissociated exciton.
}
\label{wf}
\end{figure}

\begin{figure}[htbp]
\centering
%trim=left botm right top
\begin{tabular}{l}
\includegraphics[width=0.3\textwidth]{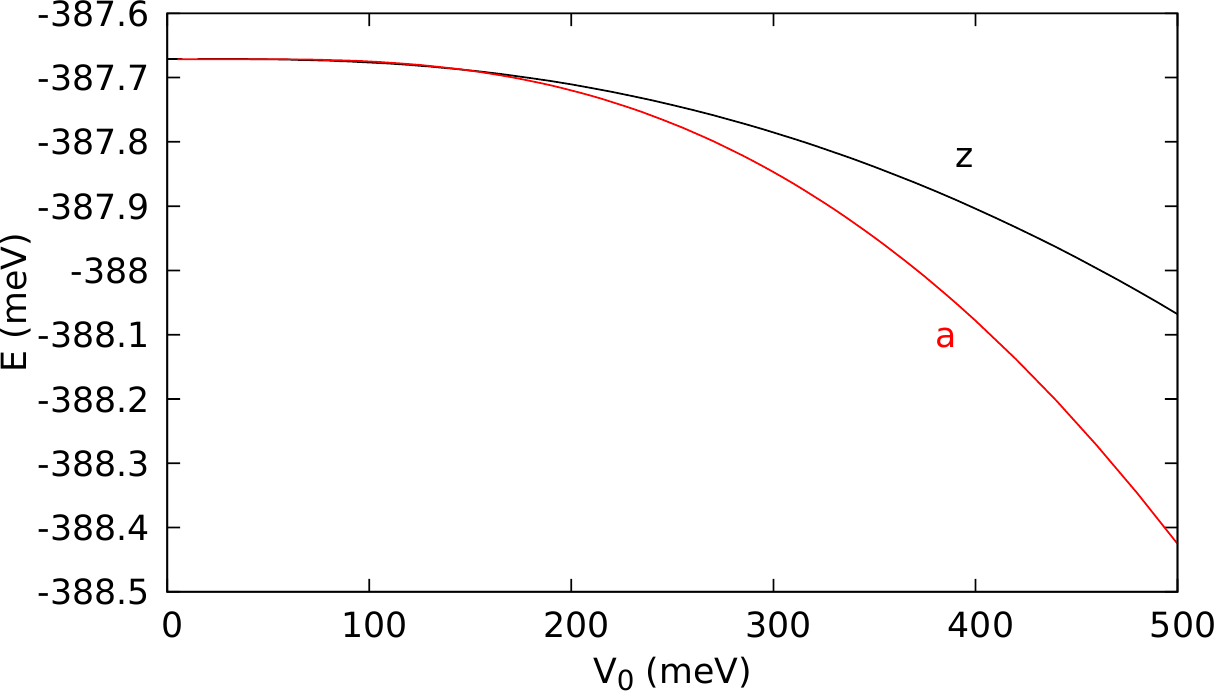} \put(-30,25){\color{black}(a)}  \\
\includegraphics[width=0.3\textwidth]{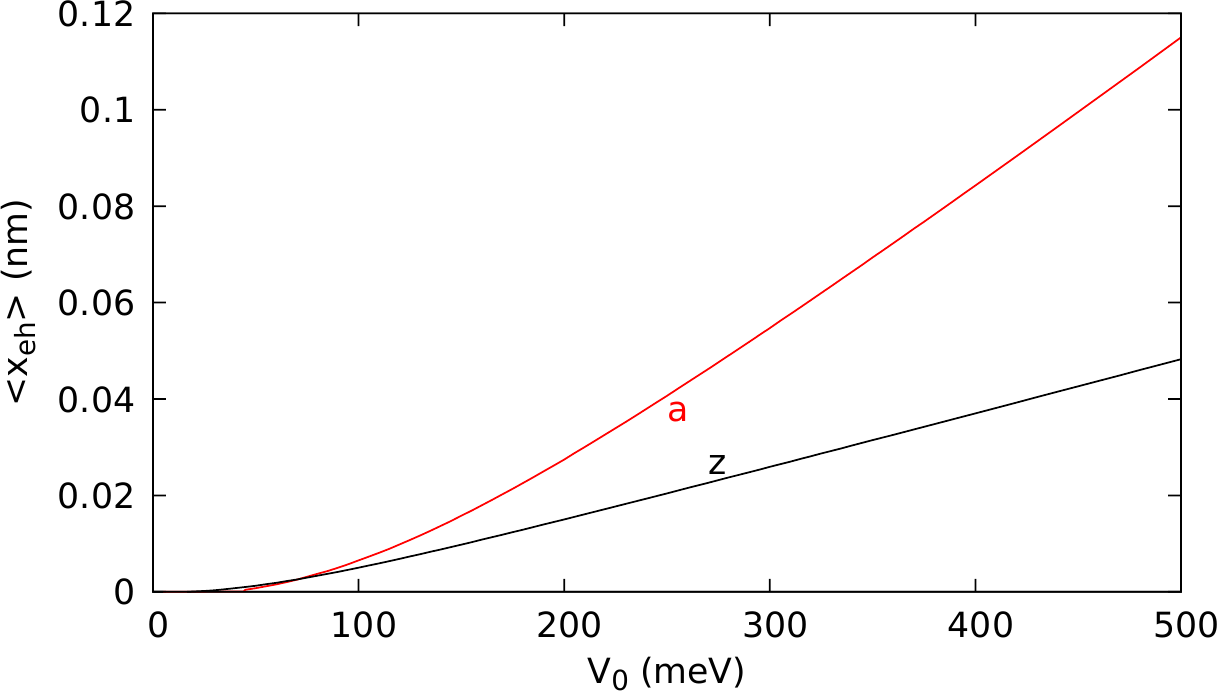} \put(-20,45){\color{black}(b)}  \\
\includegraphics[width=0.3\textwidth]{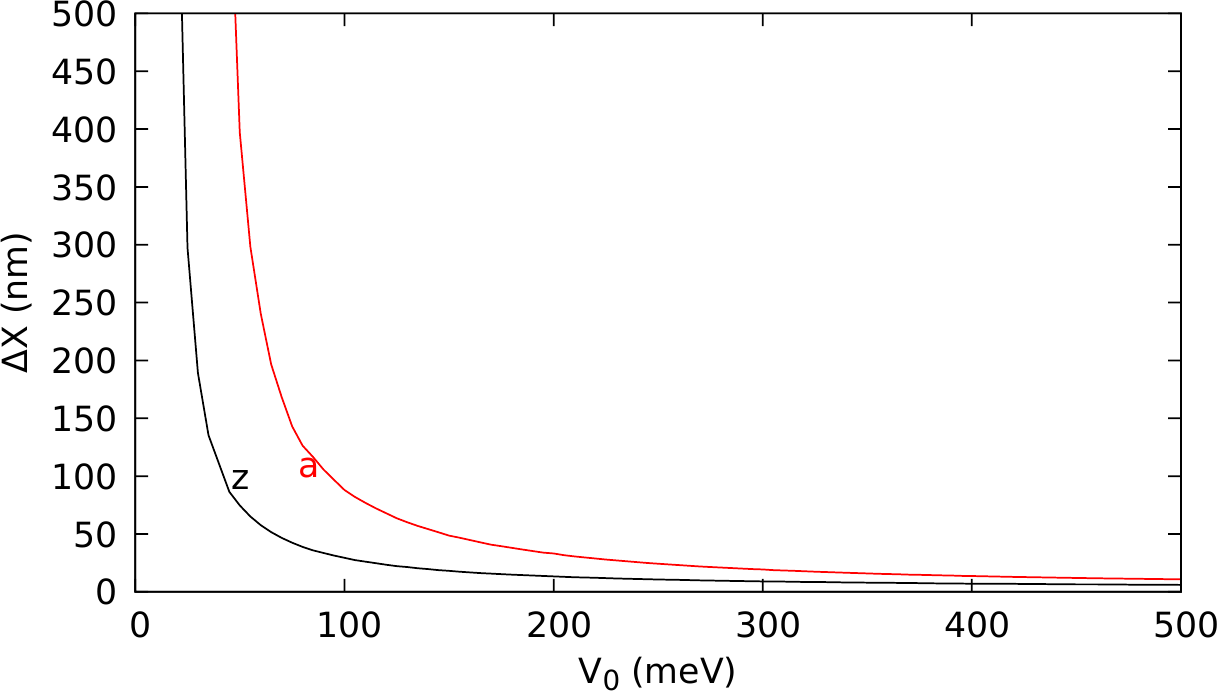} \put(-20,45){\color{black}(c)}  \\
\includegraphics[width=0.3\textwidth]{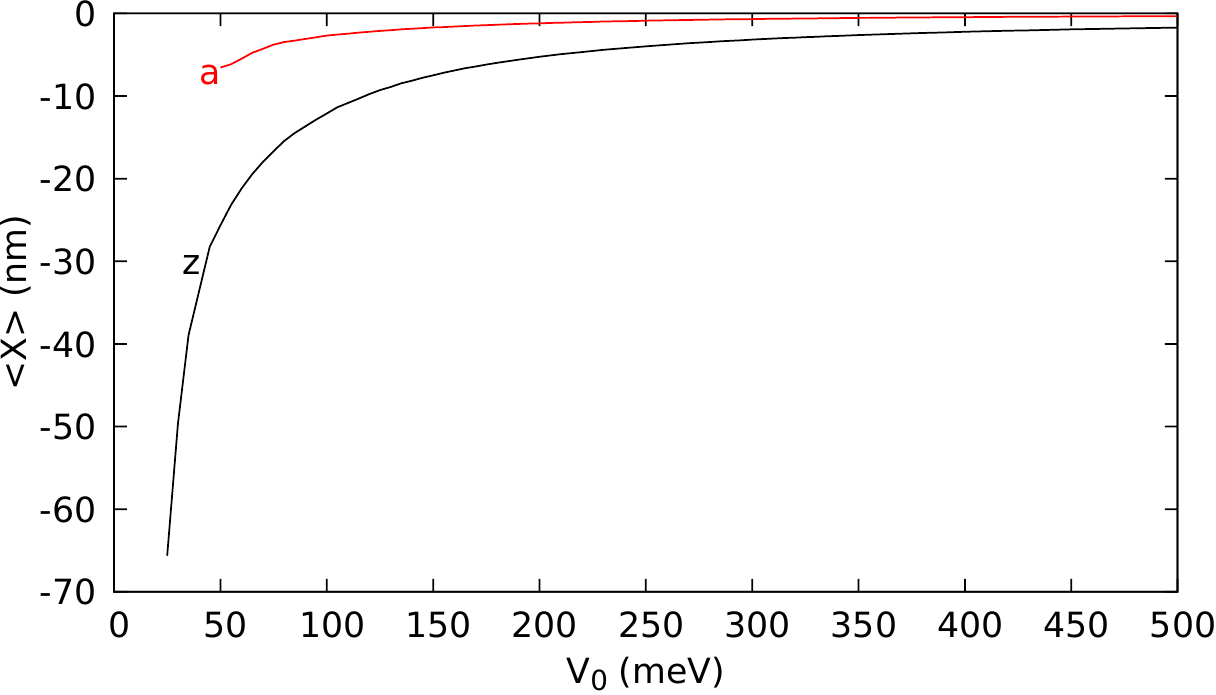} \put(-20,45){\color{black}(d)}  \\
\end{tabular}
\caption{
(a) Energy of the lowest bound exciton state as a function of $V_0$ for monolayer phosphorene with the direction defining the p-i-n junction $x$ 
aligned to the zigzag ('z', black line) or armchair ('a', red line) crystal direction (b) The average shift of the electron and hole position along the direction of the junction $x_{eh}$. (c) The exciton localization width $\Delta X$.   
(d) The average position of the exciton center of mass. %The  results presented in (a-d) correspond to the Hamiltonian ground state for $V_0<387.67/2$ meV. For larger $V_0$ the results correspond
%to the lowest-energy bound exciton state.
}
\label{wf}
\end{figure}

\begin{figure}[htbp]
\centering
%trim=left botm right top
\begin{tabular}{l}
\includegraphics[width=0.3\textwidth]{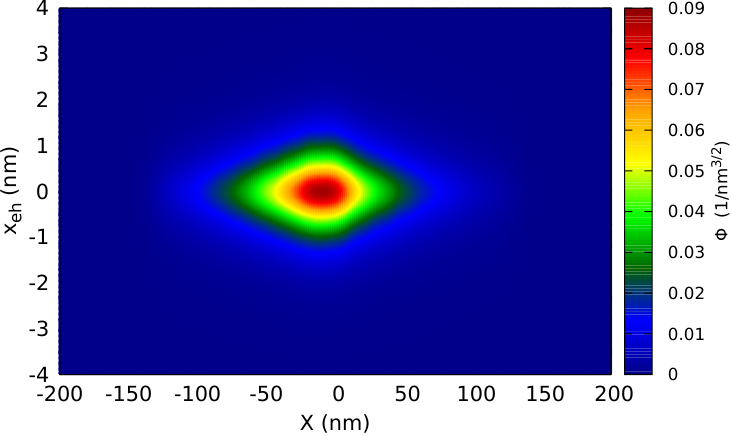} \put(-40,15){\color{white}(a)} \put(-120,15){\color{white}$z,V_0=100$ meV}  \\
\includegraphics[width=0.3\textwidth]{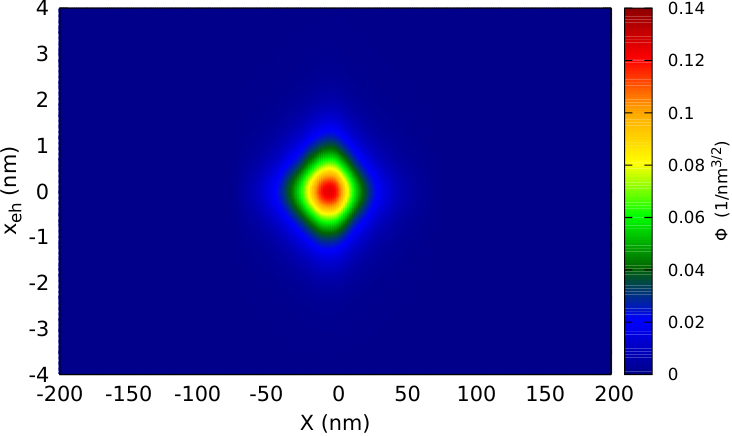} \put(-40,15){\color{white}(b)} \put(-120,15){\color{white}$z,V_0=200$ meV}  \\
\includegraphics[width=0.3\textwidth]{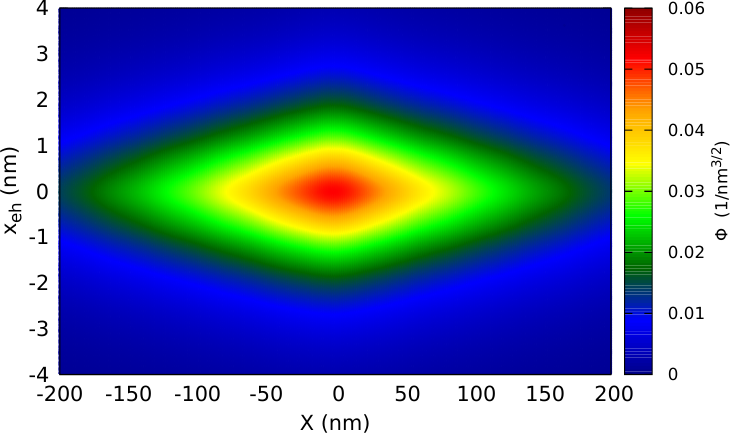} \put(-40,15){\color{white}(c)}  \put(-120,15){\color{white}$a,V_0$=100 meV} \\
\includegraphics[width=0.3\textwidth]{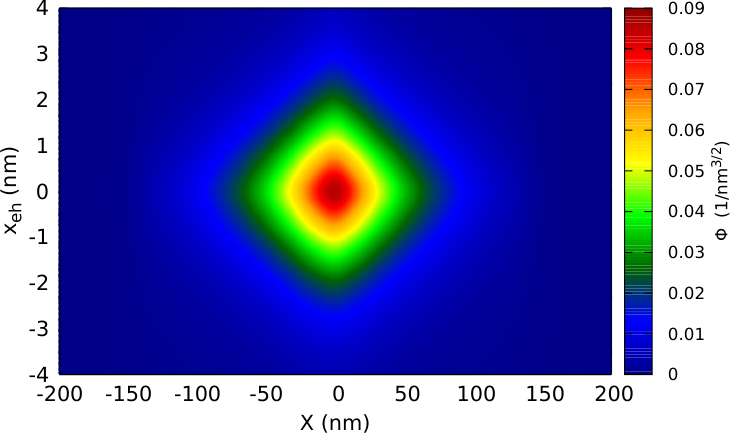} \put(-40,15){\color{white}(d)} \put(-120,15){\color{white}$a,V_0$=200 meV} \\ \\
\end{tabular}
\caption{
Cross section of the wave function taken at $y_{eh}=0$ $\Phi(x_{eh},y_{eh}=0,X)$ for $d=15$ nm and phosporene parameters.
In (a) and (b) the p-i-n junction is defined along the zigzag crystal direction. Armchair direction is chosen in (c) and (d).
% The results of (a) and (c) correspond to the ground-state and the ones in (b) and (d) to the lowest-energy bound exciton state.
}
\label{wf}
\end{figure}

\section{Results and discussion}
\subsection{MoSe$_2$}
We first consider the material where the exciton trapping at the junction was accomplished \cite{na}, MoSe$_2$, the transition-metal dichalcogenide for which
an isotropic effective mass model can be adopted 
with $m^x_e=m^y_e=0.7m_0$ \cite{na,cos} and $m^x_e=m^y_e=0.6m_0$ \cite{na,cos2}.
We take the dielectric constant $\epsilon=4.4$ and the screening length $r_0=0.886$ nm after Ref. \cite{na}.

Figure 1(b) shows the lowest bound exciton state density calculated as $\rho_X(X)=\int_{-\infty}^\infty d_{x_{eh}}\int_{-\infty}^\infty d_{y_{eh}} |\Phi(x_{eh},y_{eh},X)|^2$ for $d=15$ nm. The exciton is localized near the area where the 
potential gradient is maximal, and the exciton localization increases with $V_0$. 
The center of the exciton is localized off the center of the junction on the side of the junction 
where the heavier carrier -- here the electron -- has a lower potential energy (see below). Figure 1(c) shows the density 
over the horizontal electron-hole distance $\rho_{x_{eh}}(x_{eh})=\int_{-\infty}^\infty d_{X}\int_{-\infty}^\infty d_{y_{eh}} |\Phi(x_{eh},y_{eh},X)|^2$. The potential of the considered range produces only a small shift
of the density to a more positive $x_{eh}$, with the electron shifted to a more positive $x$ and the hole to a more negative $x$ position.
The electron and hole densities calculated from the total wave function (not shown)
are nearly identical to the exciton density $\rho_X$ due to the large extent of $\rho_X$
and the relative strong localization of the pair.

Figure 2 displays the cross section of the lowest energy bound electron-hole wave function taken at $y_{eh}=0$
for $V_0=50$ meV, 100 meV and 200 meV. The range of the localization changes radically with $V_0$ only in the $X$ coordinate and not in $x_{eh}$.

The process of exciton localization at the junction is illustrated in Fig. 3 where we considered 
several junction width parameters $d$. The energy decrease with $V_0$ in Fig. 3(a) is a signature of the carrier separation
by the electric field which, as we find, does not start at zero $V_0$ (Fig. 3(a)). 
The exciton localization in the region of a large electric field is associated with the cost of additional kinetic energy. For larger values of $d$
the electric field at the junction is lower, but the extent of  exciton localization at the junction is decreased. 
The energy shift  in the homogenous electric field falls with the square of the field \cite{starkchav,starkchav2} which for the present potential implies a $-1/d^2$ dependence  while the exciton confinement energy depends on the localization as $1/d^2$. The two effects seem to cancel
each other out in the constant energy range of small $V_0$.
We observe that a nonzero value of 
the potential step is necessary to induce a significant dipole moment (Fig. 3(b)). 
As non-zero $V_0$ is introduced the range of exciton localization (Fig. 3(c)) becomes finite.
However, for low $V_0$ most of the exciton stays outside of the action range 
of the local electric field (potential gradient) at the junction. For all studied $d$ values, the exciton size in terms of $\Delta X=\langle (X-\langle X \rangle )^2\rangle^{1/2}$ becomes equal to double the width parameter $\Delta X=2d$ 
 for $V_0\simeq 100$ meV.
Only once the exciton
is localized at the junction does the electron-hole system energy start to decrease visibly.

For comparison in Fig. 3(a) and (b) we plotted with the dashed red line the results for the homogenous electric field, i.e.
for the potential of the junction replaced by its linear approximation: $V_l(x)=-e\frac{2V_0}{\pi}\frac{x}{d}$,
for $d=30$ nm. For $V_l(x)$ potential the dipole moment grows linearly with $V_0$ [Fig. 3(b)] starting from $V_0=0$. The linear dependence
of the dipole moment via its interaction with the field produces the parabolic $E(V_0)$ dependence
(the dashed line in Fig. 3(a)). For the junction potential $V(x)$,
the growth of the dipole moment with $V_0$ follows after a delay. For all $d$ considered
the dipole moment starts to grow for $V_0\gtrsim 20$ meV and becomes a linear function of $V_0$ 
only above $\simeq$ 100 meV. 

Figure 3(d) shows that the center of mass of the exciton approaches the center of the junction only in the $V_0\rightarrow \infty$ limit, and for finite $V_0$ it stays on the positive side of the center of the junction.
We return to this point for phosphorene, where the effect is more pronounced.

%For $V_0=100$ meV the average electron-hole separation in the $x$ direction $\langle x_{eh}\rangle$ (see Fig. 3(b))  is equal 
%to $0.05124$ nm, $0.016916$ nm, 0.008449 nm and $0.004226$ nm for $d=5$ nm, 15nm, and 30 nm, respectively.
%The dipole moment is linear in $1/d$, i.e. the product $\langle x_{eh} \rangle d \simeq 0.25$ nm$^2$ for all the considered values of the junction width $d$. 

%The dipole moment
%is larger for shorter junctions (Fig. 3(b)). 	Figure 3(b) shows that the average electron-hole separation $\langle x_{eh}\rangle$  rises  from very low values 
%to a steeper dependence on $V_0$ only for $V_0>\sim 20$ meV and the starting point of this increase on the $V_0$ scale does not have a noticeable dependence on $d$.  

The ground-state energy $E_{gs}(0)=-216.89$ meV obtained at $V_0$ (Ref. \cite{na} indicates -217 meV) implies that
the dissociated electron-hole pair appears in the ground state for $V_0\geq E_{gs}/2$. 
For $V_0/2$ above $E_{gs}$ the bound hole-pair is not the ground state of the system,
 similarly to the studies of the Stark effect for a nonzero homogeneous electric field  \cite{na,starkchav,starkchav2}. 
 The applied basis resolves two separate minima for the bound and dissociated exciton. The results of the Hamiltonian
diagonalization as a function of $\Delta_{x_{eh}}=\Delta_{y_{eh}}$ 
are given in Fig. 4, with the lower energy state corresponding to the bound exciton or the dissociated pair
for $V_0=100$ meV and 300 meV, respectively.
Here, we discuss only the properties of the bound electron-hole state  for both $V_0$ below or above
the dissociation  threshold. The gradient minimalization method applied for optimization of the non-linear variational parameters tends to the closest energy minimum, 
so one can choose between the dissociated or bound exciton states.

\subsection{phosphorene}

The  effective masses in phosphorene exhibit  strong anisotropy with larger (smaller) values
in the zigzag (armchair) crystal direction \cite{jakies,34}. 
For monolayer black phosphorous, we adopted the values of effective masses
derived by fitting the results of the single-band approximation to the 
results of the atomistic tight-binding \cite{tb} modelling for the harmonic oscillator confinement
potential of Ref. \cite{szafran}. That is, we take $m^e_a=0.17037m_0$, $m^e_z=0.85327m_0$
for the electron masses in the armchair and zigzag directions \cite{szafran}, respectively.
For hole we take $m^h_a=0.18972 m_0$ for the armchair direction \cite{szafran}.  
The hole mass in the zigzag direction was not adopted from Ref. \cite{szafran} but set to $2.8 m_0$.
The literature values for the zigzag mass of the hole vary from $\simeq 1.13m_0$ \cite{34,43}
to $\simeq 5$ \cite{48} or $\simeq 6$ \cite{47}. The selected choice of the effective masses provides the electron-hole pair reduced
masses that agree with the ones given in Table 2 of Ref. \cite{castelano} for monolayer phosphorene.
For the interaction potential we take the screening length of $r_0=1.079$ nm 
and the effective dielectric constant of $\epsilon=2.4$ after Supporting Information to Ref. \cite{castelano}.

In calculations for phosphorene we fix the junction width parameter to $d=15$ nm. 
We keep the junction potential variation along the $x$ axis that we align 
with either armchair or zigzag crystal direction. The reaction of the exciton energy to the external potential
is prompter for the junction defined along the armchair direction (Fig. 5(a)), and the induced dipole moment
[Fig. 5(b)] is also larger in this direction where the masses are lighter. 
This finding is consistent with the conclusions of Ref. \cite{starkchav} for the case of a homogeneous electric field 
in phosphorene.

The cross sections of the wave function for $y_{eh}=0$ is plotted in Fig. 6 for the zigzag (a,b) and armchair (c,d) orientation
of the wave function. The orientation of the junction has a pronounced effect on the exciton localization along the $X$ axis
with the total mass $M_x=0.36m_0$ for the armchair orientation $M_X=3.65m_0$ for the junction defined along the zigzag crystal direction.
The reduced mass in the $x_{eh}$ direction is $0.089m_0$ for the armchair and $0.65m_0$ for the zigzag orientation, hence
the varied localization along the vertical axis of Fig. 6. 
Fig. 5(c) indicates that the width of the exciton wave
function becomes equal to $2d$ for $V_0\simeq 100$ meV ($V_0\simeq 210$ meV) for the junction defined along the zigzag (armchair) direction.

For phosphorene, the center of the exciton density is shifted to the left
of the junction center [Fig. 6 and Fig. 5(d)]. The shift is smaller for the armchair orientation of the junction [Fig. 6(c,d)]
 and  large for the zigzag-oriented junction [Fig. 6(a,b) and Fig. 5(d)]. 
As a general rule, the shift is large when one of the carriers is much heavier than the other and the shift 
appears in the direction of lower potential energy in the $V(x)$ potential for the heavier carrier.
In the results for MoSe$_2$ (precedent subsection), the shift was observed to the positive side of the junction and the electron
was slightly heavier. For phosphorene, the hole is heavier and the center of mass is shifted to the negative side of the junction.
For the zigzag orientation, the mass difference is very pronounced; hence, the strong shift, much larger than the average electron-hole distance.
The shift is directly related to the potential landscape at the junction. In Fig. 7 we plotted the external potential
for both carriers, i.e. $V(x_e)-V(x_h)$ on the ($X$, $x_{eh})$ plane.
% The arguments of the $V(x_{e/h})$
 %functions in terms of the center of mass are expressed as $x_e=X+\frac{m_h}{M}{x_{eh}}$ and $x_h=X-\frac{m_e}{M}{x_{eh}}$. 	
The anisotropy of the masses translates to a shift of the potential minimum to the negative side of $X$, which produces 
the pronounced asymmetry for the zigzag orientation of the junction. The interaction potential of Hamiltonian (3) is independent of $X$ 
and creates a valley along $x_{eh}=0$ of depth that depends on $y_{eh}$. The displacement of the wave function to positive $x_{eh}$ 
is not very pronounced (see Fig. 6 and Fig. 5(b)) since the strong interaction potential keeps the carriers close to one another. Instead, the wave function
is shifted to the negative side of the junction where the minimum of the potential landscape is closer to the $x_{eh}=0$ axis.
The effect is particularly well seen in Fig. 6(a,b) (see also Fig. 5(d)).
Note that for a homogeneous electric field, only the reduced masses and not the separate masses of the electron and hole matter for the properties of the bound exciton states \cite{starkchav}.

\begin{figure}[htbp]
\centering
%trim=left botm right top
\begin{tabular}{l}
\includegraphics[width=0.3\textwidth]{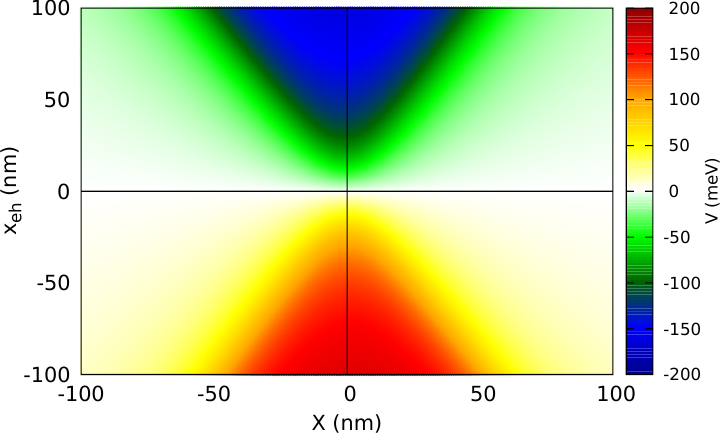} \put(-40,20){(a)} \put(-130,20){$a$}  \\
\includegraphics[width=0.3\textwidth]{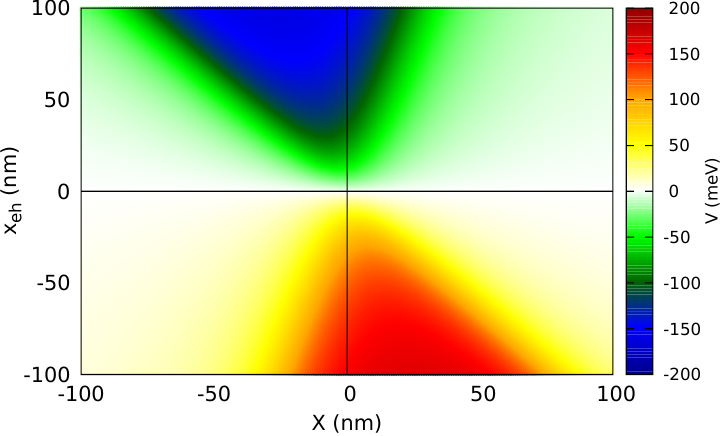} \put(-40,20){(b)} \put(-130,20){$z$}  \\
\end{tabular}
\caption{Potential landscape $V(x_e)-V(x_h)$ at the junction for $d=15$ nm, $V=100$ meV for phosphorene parameters
and the junction defined along the armchair (a) and zigzag (b) crystal direction. }
\label{wf}
\end{figure}

\section{Summary and Conclusions}
We have studied bound electron-hole pairs at a p-i-n junction 
defined within a two-dimensional crystal using a model with a finite spatial range 
and potential step $V_0$ across the junction with the electric field in the center of the junction proportional to $V_0$.
The problem was solved with a  variational calculation using the basis of Gaussians
centered on a grid spanned by the relative coordinates of the carriers and the center of mass, which
is not separable in the inhomogeneous electric field.
The appearance of the dipole moment induced by the local electric field at the junction
is accompanied by a build-up of the exciton kinetic energy as a result of the localization at the junction.
The compensation of the two contributions leads to a range of values of the potential step of the junction $V_0$
that has a negligible influence on the exciton energy.
For MoSe$_2$ and phosphorene, localization of the exciton wave function
in a range comparable with the junction length requires a potential difference across the junction
of the order of $V_0\simeq 100$ meV. For lower $V_0$ values, the carriers occupy the region in space 
where the electric field is much smaller than at the center of the junction.
The induced dipole moment dependence on the potential step $V_0$ deviates from linear, which is expected
for the Stark effect in a homogenous electric field. In consequence, the exciton energy is not a parabolic function of $V_0$.
We have demonstrated that the localized exciton is shifted off the center of the junction in the direction in which
the junction potential energy for the heavier carrier is lower.

\section*{Acknowledgments}
Calculations for this work were performed on the PL-GRID infrastructure.

% jak dochodzi do dysocjacji: przez pojawienie sie innego stanu, nie ma gladkiego przejscia - tutaj podobnie jak w teoriach dotyczacych efektu Starka w zakresie pól, zbliżonych do dysocjacji rozmiar ekscytonu (jednak narysować) jest znacznie mniejszy od złącza. On tam się mieści w całości. 

\end{document}